\documentclass{article}
\usepackage{amsmath,graphicx,epstopdf}

\begin{document}

\title{Self Organized Criticality in an one dimensional magnetized grid. Application to GRB X-ray afterglows}

\author{Tiberiu Harko\footnote{Department of Mathematics, University College London, Gower Street, London
WC1E 6BT, United Kingdom, E-mail:
t.harko@ucl.ac.uk}, Gabriela Mocanu\footnote{Astronomical Observatory Cluj-Napoca, Astronomical Institute of the Romanian Academy, Cire\c{s}ilor Street no. 15, 400487 Cluj-Napoca, Romania, E-mail: gabriela.mocanu@ubbcluj.ro}, Nicoleta Stroia\footnote{Technical University, Department of Automation, C-tin Daicoviciu Street no. 15
400020, Cluj-Napoca, Romania, E-mail: nstroia@mail.utcluj.ro}}
\date{}
\maketitle

\begin{abstract}
A simplified one dimensional grid is used to model the evolution of magnetized plasma flow. We implement diffusion laws similar to those so-far used to model magnetic reconnection with Cellular Automata. As a novelty, we also explicitly superimpose a background flow. The aim is to numerically investigate the possibility that Self-Organized Criticality appears in a one dimensional magnetized flow. The cellular automaton's cells store information about the parameter relevant to the evolution of the system being modelled. Under the assumption that this parameter stands for the magnetic field, the magnetic energy released by one volume during one individual relaxation event is also computed. Our results show that indeed in this system Self-Organized Criticality is established. The possible applications of this model to the study of the X-ray afterglows of GRBs is also briefly considered.
\end{abstract}


\section{Introduction}

GRBs are explosive cosmic gamma-ray emissions, with typical energy fluxes of the order of $10^{-8}$ to $ 5 \times 10^{-7}$ J m${}^{-2}$, and durations that range from $10^{-2}$ to  $10^{3}$ s \cite{Pi92}. Since their distribution is isotropic, they are believed to have a cosmological origin, being located at extra-galactic distances \cite{Pi92}. The observable effects of GRBs are produced by the dissipation of kinetic energy from a relativistically expanding plasma, though the underlying progenitor model is, as yet, unknown. Reviews and discussion of new models may be found in, e.g., \cite{BaPaSp87+,Ha03,ToWuMe09,SuIo10,Be10,Re11,VuBePo11,ZhYa10,Zhetal13,Ha14}.

The detection of X-ray flares in nearly half of the observed GRB afterglows \cite{Chin, Mar} has added a new mystery to the GRB puzzle. The present astrophysical observations suggest that the central engines of bursts, after the gamma-ray emission has ended, still have long periods of
activity, during which energetic explosions eject relativistic materials, leading to late-time X-ray emission \cite{wang2013}. The GRB observations show that strong magnetic fields play a key role during the acceleration of the outflow and the prompt emission \cite{Lyut}, while any remaining magnetization at large distance can affect the interactions of the flow with the external medium.

A statistical analysis of the X-ray flares of GRBs with known redshifts was performed in \cite{wang2013}. It was shown that X-ray flares and Solar flares share three statistical properties: a) power-law energy frequency distributions, b) power-law duration-time frequency distributions, and c) power-law waiting time distributions, respectively. All these distributions can be well understood within the physical framework of a magnetic reconnection-driven {\it Self-Organized Criticality system}. These statistical similarities, together with the fact that Solar flares are triggered by a magnetic reconnection process taking place in the atmosphere of the Sun, suggest that X-ray flares originate from magnetic reconnection-driven events possibly involved in ultra-strongly magnetized millisecond pulsars \cite{wang2013}.

In \cite{Dai} and \cite{Metz} it was already proposed that GRB X-ray flares may be powered by magnetic reconnection events. Magnetic reconnection may occur near the photosphere if the outflow develops an alternating field structure due to e.g. magnetic instabilities or a misalignment between the magnetic and rotation axes. The magnetic dissipation model is favoured by observations \cite{Metz}.

\cite{bak1988} proposed that the numerous spatial extended systems, which exhibit a number of properties which may be shortly characterized as flicker noise for the temporal evolution and self-similar (fractal) behavior for the spatial evolution can be organized under the same principle of Nature, namely as systems in the state of {\it Self Organized Criticality} (SOC) \cite{Sornette}.

There is a lot of literature covering the aspects of SOC in astrophysics~(see, e.g., \cite{aschSoc2013} and references therein), ranging from simple but effective numerical models~\cite{bak1988,lu1991,moc2012}, to sophisticated analytical models as in~\cite{lu1995} and \cite{gilSor1996}. The possible areas of physics in which these models prove to reproduce observational data is very wide. This is a consequence of the fact that SOC may be an underlying principle in Nature, as the initial intuition of~\cite{bak1988} stated. Examples in astrophysics include organization of plasma in accretion discs~\cite{mineshige1994} to explain observational data such as those discussed in~\cite{mocanuAN2012}, axion clouds around black hole~\cite{moc2012}, solar flares~\cite{dimitropoulou2011,isliker1998,isliker2002,lu1991}.

\cite{StSv1996} have analyzed GRB X-Ray data, and they did find that all observational results can be fitted into a framework based on a stochastic pulse avalanche model running in a near-critical regime. Their reasoning is reproduced here: 1. All GRBs can be described as different random realizations of the same simply organized stochastic process within narrow ranges of the parameters of the process. 2. The stochastic process should be scale invariant in time. 3. The stochastic process works near its critical regime. This would explain the large morphological diversity of GRBs. They propose that the flaring events are based on magnetic reconnection, but their simulations do not include evolution equations for magnetic fields. While this type of statistics of data can be explained by turbulence, the analysis performed by~\cite{wang2013} is clearly in favour of a Self Organized Criticality model.

It is the goal of the present paper to investigate from the point of view of SOC one dimensional magnetized systems undergoing slow driving. More specifically, this study aims at describing a specific behaviour, i.e., that of magnetic field on a space scale that is global. Thus we do not attempt to solve the MHD equations nor do we even take all of them into account. We focus on the magnetic field as a being a relevant parameter, since it is a field that couples to the radiation field and since radiation is our main source of information on astrophysical sources.  Also, the magnetic field is a parameter that exhibits a transition in its behaviour and this transition can be explained based on the Laplacian of the magnetic field reaching a critical value. The evolution equation for this parameter is the magnetic induction equation, which describes the time behaviour of a magnetic field in a resistive plasma medium in which the MHD approximation holds~\cite{Priest2000}, with a background flow.

Magnetic reconnection was investigated independently of the idea of SOC as a natural phenomenon occurring in all branches of astrophysics~\cite{Parker,Sweet}. It can be viewed as a change in the magnetic field topology when the value of a certain parameter locally reaches a threshold value. The change in topology occurs fundamentally because the configuration is no longer of minimum energy. A part or all of the excess energy is lost to radiation. Extension of the idea of magnetic reconnection to SOC is nothing more but moving from the local scale of the reconnection site to the entire system. So conceptually SOC in a magnetized environment is expected and it is thought to be the answer to many unanswered questions in astrophysics~\cite{Zhang08,AschOwn2013}. This expectation was confirmed by observational data of solar flares (a review of theory and data is presented in, e.g., \cite{Charbonneau}).

The present paper is organized as follows. In Section~\ref{sect:SOC} we introduce the magnetic induction equation for plasma flows, and we present its one-dimensional version. The discretization of the induction equation as well as the simulation procedure is presented in Section~\ref{disc}. The results of our simulations for one dimensional magnetic flows and the establishment of the SOC are presented in Section~\ref{sect:sims}.  The applications of the model for the study of GRB afterglows are briefly presented in Section~\ref{grb}. We discuss and conclude our results in Sections~\ref{discussion} and~\ref{concl}.

\section{Self organized criticality in magnetized flows}\label{sect:SOC}

The aim of this section is to investigate whether or not SOC is possible in a one dimensional magnetic field set-up.

\subsection{Preliminaries: The theoretical model}

The set of MagnetoHydrodynamic equations which (under some assumptions regarding the ratio of microscopic and macroscopic time-and-space scales) analytically describe general magnetized flows are inhomogeneous nonlinear partial-differential equations, which are sometimes anisotropic. Numerical techniques to solve this set of equations have evolved, but one still cannot solve them in full generality not even locally.

Cellular Automata (CA) simulations are a valid alternative to the full numerical treatment when only qualitative and quantitative answers of a specific kind are sought for. In such a setup the velocity profile has not been consistently included, to our knowledge. In fact, most of the times it has been considered nonexistent and the convective part was replaced by a stochastic loading phase. The following treatment aims at a middle ground between ignoring the flow completely and the exact numerical solution of the equation (not accessible to a CA of this type).

The skeleton of our approach to calculating the stress, number of events and released energy has been used in the literature, see e.g.~\cite{lu1991}; the novelty in this work is explicitly taking into account the background flow characteristic to GRBs.

Thus we consider the induction equation
\begin{equation}
\frac{\partial \vec{B}}{\partial t} = \nabla \times \left ( \vec{v} \times \vec{B} \right ) + \eta \nabla ^2 \vec{B},
\end{equation}
and define the control parameter as
\begin{equation}
\vec{G} = -\frac{1}{nn}\nabla ^2 \vec{B},
\end{equation}
where $\vec{B}$ is the magnetic field, $\vec{v}$ is the flow velocity, $\eta$ is the constant plasma resistivity and the notation $nn$ denotes the number of nearest neighbours in the configuration.

The set-up consists of a flux tube aligned with the $Oz$ axis of a system of coordinates fixed at the footpoint of the tube. A magnetic field is present, $\vec{B}=(B(z),0,0)$ and we allow for a background flow $\vec{v}=(0,0,v(z))$. We emphasize that the setup is thus two dimensional. In this setup, with $nn=2$, the magnetic induction equation and the control parameter become
\begin{equation}
\frac{\partial B}{\partial t} = -\frac{\partial (Bv)}{\partial z} + \eta \frac{\partial ^2 B}{\partial z^2}, \quad G = -\frac{1}{2} \frac{\partial ^2 B}{\partial z^2} .
\end{equation}

In astrophysical conditions the classical resistivity is very small and the magnetic field behaves macroscopically as if the diffusion term in the magnetic induction equation would be zero. This behaviour is dictated by the magnetic Reynolds number, $R_m = UL/\eta$ where $U$ and $L$ are characteristic velocity and space-scale respectively. However, under certain conditions, in relatively small volumes, the diffusive behaviour becomes dominant and the magnetic field lines reconnect. To model this in our grid, we shall assume that the physics occurs in two different regimes and the switch between these two regimes is dictated by the behaviour of the control parameter.  The control parameter tells us what is the value of the difference between the magnetic field between one grid point and its neighbours. By definition this control parameter is thus local, and its characteristic scale $l$ is very small. If $L$ is the characteristic length scale of the simulation, then in our case $l/L$ is at least $0.002$. If the control parameter exceeds a certain threshold, then what happens locally becomes worthwhile inspecting. Since we may assume the velocity does not change in order of magnitude, and since the diffusivity $\eta$ is constant (the plasma does not change), the ratio between the macroscopic Reynolds number and that of the local Reynolds number is of the same order of $l/L$. This can be viewed as a reason why the control parameter changes the behaviour of the Reynolds number. To summarise, the convective behaviour is the main framework. If the control parameter becomes critical the magnetic field evolution is given, for a brief period of time, by a diffusive behaviour. Once this local criticality is relaxed, control is given back to the convective behaviour.

Stated more clearly, the induction equation is
\begin{equation}\label{Rm}
\frac{\partial B}{\partial t} = \left \{ \begin{array}{rl}  -\frac{\partial (Bv)}{\partial z} & \text{ high } R_m>>1,  \\ \eta \frac{\partial ^2 B}{\partial z^2} & \text{ low } R_m<<1.\\
\end{array}\right .
\end{equation}

We denote the term $- v \frac{\partial B}{\partial z}$ in \eqref{Rm} by $S(z,t)$ and consider it as a stochastic source term.

Equations~\eqref{Rm} are brought to dimensionless form by the transformations $t=\alpha T$, $B=bB_{0}$, $z = \beta Z$, $v = v_0 V$, $G=gB_{0}/\beta ^2$, where $\alpha$ is a characteristic timescale, $B_{0}$ is the characteristic magnetic field, $\beta$ is the a characteristic length scale defined here as the distance travelled in time $\alpha$ by a perturbation moving with Alfven velocity  $v_A$ in a medium with a characteristic magnetic field equal to $B_{0}$, and $v_0$ is the initial velocity of the flow.

With these parameters, the diffusive description becomes
\begin{equation}
\frac{\partial b}{\partial T} = k \frac{\partial ^2 b}{\partial Z^2}, \quad k = \frac{\alpha \eta}{\beta ^2},
\end{equation}
while the convective description becomes
\begin{equation}
\frac{\partial b}{\partial T} = - \chi b \frac{\partial V}{\partial Z} + s(Z,T), \quad \chi = \frac{\alpha v_0}{\beta},
\end{equation}
where $s(Z,T) = \alpha  S(z,t) /B_0$; the dimensionless expression for the control parameter is
\begin{equation}
g = - \frac{1}{2}\frac{\partial ^2 b}{\partial Z^2}.
\end{equation}

\section{Discretization of the induction equation}\label{disc}

The discretisation of the induction equation follows the general rules
\begin{equation}
\frac{b_{i,j+1}-b_{i,j}}{\Delta T} = \frac{k}{(\Delta Z)^2} \left [ b_{i+1,j} + b_{i-1,j} - 2b_{i,j} \right ],
\end{equation}
\begin{equation}
\frac{b_{i,j+1}-b_{i,j}}{\Delta T} = - \chi b_{i,j}\frac{V_{i+1,j}-V_{i,j}}{\Delta Z} + s_{i,j},
\end{equation}
\begin{equation}
g_{i,j} = -\frac{1}{2\left ( \Delta Z \right )^2} \left [ b_{i+1,j} + b_{i-1,j} - 2b_{i,j} \right ], \label{eq:stress}
\end{equation}
where the notation $f_{i,j}$ stands for the value of the parameter $f(Z,T)$ evaluated at $Z=i\Delta Z$ and $T=j\Delta T$.

For the stochastic source term, at each time step $j$, we choose to update the magnetic field at a randomly chosen position $k_j$ as follows
\begin{equation}\label{eq:loading}
s_{i,j} = b_{i,j}(1+\Delta T \epsilon) \delta _{i,k_j},
\end{equation}
where $\epsilon$ is some positive small number $< 1$.

The diffusion description, written in update form and using Eq.~(\ref{eq:stress}) becomes
\begin{equation}
b_{i,j+1} = b_{i,j} - \zeta g_{i,j}, \quad \zeta = 2 k \Delta T.
\end{equation}

We will consider flows with a velocity decreasing as the spatial grid index increases, such that $V_{i+1,j}<V_{i,j}$, and, even more, propagating flows such that the velocity is zero for points not yet reached by the wavefront, i.e., $V_{i+1,j}-V_{i,j} = -V_{i,j}$. With this assumption and written in update form, the convection is described as
\begin{equation}
b_{i,j+1} = b_{i,j} \left [ 1 + V_{i,j} \theta \right ] + \Delta T s_{i,j}, \quad \theta = \chi \frac{\Delta T }{\Delta Z}.\label{eq:flow}
\end{equation}

When the critical threshold has been reached in a point of the grid, $|g_{i,j}|\geq g_{cr}$, the critical quantity is redistributed among the neighbours of this point. This corresponds to the diffusive behaviour of the induction equation,
\begin{equation}
b_{i,j+1} \to b_{i,j} - \frac{2}{3} g_{i,j},\label{eq:bi}
\end{equation}
which fixes $\zeta = 2/3$, and leads to a redistribution
\begin{equation}
b_{i\pm 1,j+1} \to b_{i\pm 1,j} + \frac{1}{3}g_{i,j}.\label{eq:bneigh}
\end{equation}

The energy released by one volume during one individual relaxation event is the magnetic energy lost in that volume
\begin{eqnarray}
E_R& =& \frac{1}{2\mu _0}\int _V \left (B_{in}^2-B_{fin}^2 \right )dV = \nonumber\\
&&\frac{1}{2\mu _0}\int _{dz} \left [ \int _{dxdy} \left (B_{in}^2-B_{fin}^2 \right )dxdy \right ] dz\label{eq:magnPress}.
\end{eqnarray}

Since the magnetic energy is not dependent on the $x$ and $y$ coordinates, and the $x$ and $y$ dimensions are negligible with respect to the $z$ dimension, we take the result of the integral with respect to these variable to be $1$. Rearranging and writing in dimensionless form, Eq.~(\ref{eq:magnPress}) becomes
\begin{equation}\label{eqen1}
e_R = \int _{dZ} \left (b_{in}^2-b_{fin}^2 \right ) dZ,
\end{equation}
where $E_R = e_R\frac{\beta B_{0}^2}{2\mu _0}$ (multiplied by unit area) and $e_R$ is a dimensionless quantity.

We approximate Eq.~(\ref{eqen1}) with its discretized counterpart, for one cell with $\Delta Z = 1$, we drop the index $R$, and use the same $i,j$ notation, thus obtaining
\begin{equation}
e_{i,j*} = b_{i,j*}^2-b_{i,j*+1}^2.
\end{equation}
This is the dimensionless energy released by one unstable cell $i$. The star on $j$ is adopted to make the following thing clear: within one time step $j$ of the simulation, the same cell, due to next neighbour interaction might become unstable more than once. The $j*$ is a subdivision of the simulation time step and it is nonzero as long as the cell is unstable. The quantity to be compared with observations, the energy emitted by the entire grid during one time step $j$ is
\begin{eqnarray}
e_j &=& \sum_{\text{unstable nodes $i$}} e_{i,j*} = \nonumber\\
&&\sum_{\text{unstable nodes $i$}} \left ( \frac{4}{3}b_{i,j*}g_{i,j*} - \frac{4}{9}g_{i,j*}^2\right ) =\nonumber\\
&& N_j \left ( \frac{4}{3}b_{i,j*}g_{i,j*} - \frac{4}{9}g_{i,j*}^2\right ),
\end{eqnarray}
where $N_j$ is the number of events needed to relax the grid during time step $j$.

\subsection{Simulations}

The procedure is summarized below
\begin{enumerate}
\item{}Initialization: a one dimensional grid with $N_Z$ grid points is initialized to hold in each cell a value for the magnetic field, $b_0$; the initial velocity ($V_{0,0}$) is some multiple of the characteristic Alfven speed for the configuration
\item{}Evolution: For each $j$ in the interval $\overline{1,N_T}$ time steps, the evolution of the system is implemented as
	\begin{itemize}
	\item{}choose a random number $k_j$ in the interval $\overline{1,N_Z}$ and update the value of the $k_j^{th}$ cell in agreement with Eq.~(\ref{eq:loading}) (stochastic loading).
	\item{}since the upward flow with velocity $V$ is deterministic, one can formally determine what cell $i_j$ the flow has reached at the current time step $j$. The numerical value in this cell 	is updated according to the first term on the left hand side of Eq.~(\ref{eq:flow}) (convective behaviour).
	\item{}at this point the mean value of the magnetic field in the grid is calculated; each component of the grid is then scaled with respect to this mean value. The critical parameter $g_{cr}$ (the difference between neighbouring cells needed to trigger a relaxation) is taken as $10\%$ of the mean value.
	\item{}\textit{for} each cell in the grid, i.e., $i$ in the interval $\overline{1,N_Z}$, $g_{i,j}$ is calculated with Eq.~(\ref{eq:stress}). \textit{If} the absolute value of the control parameter is larger than $g_{cr}$, then a diffusion behaviour is implemented in line with equations~(\ref{eq:bi}) and (\ref{eq:bneigh}). This sweep of the entire spatial grid is done \textit{while} cells with critical parameter can still be found. A variable $N_j$ stores how many times (for one particular time step $j$) cells in the grid were relaxed.	
	\end{itemize}
\item{}Results: $N_j$ represents the number of events needed at each time step $j$ in order to fully relax the grid. Each of these numbers is what is also called the avalanche size and the vector $N$ is used to produce the event size distribution.
\end{enumerate}

\section{Self organized criticality in one dimensional magnetic flows}\label{sect:sims}

We discuss a set of simulations we performed, with parameters $N_T=10^5$, $N_Z=500$, $b_0=1$, $\epsilon = 0.3$, $\Delta T = \Delta Z = 1$, $\chi = \theta$ and in which we varied the profile of $V_{i,j}$ according to the laws $V_{i,j} = {\rm constant}$, $V_{i,j} = \sqrt{j^{-1}}$, $V_{i,j} = j^{-1}$ and for various values of $\chi$. For comparison purposes the results in which convection is ignored are also included (marked with $V=0$).
SOC occurs when cells in the system have a value for the parameter larger or equal to the critical parameter \textit{and} when this state is spread out in the simulation grid with no preferred length scale. The lack of preference is seen in the power-law shape of the event size distribution $D(N)$. $D(N)$ is the number of times during the simulation in which a number $N$ of events was needed to relax the grid. A fit of the type $D(N)\sim N^{a}$ consistently gives a value of $a$ close to $1$ for all the parameter combinations shown in Table~\ref{tab:params}, as it is expected theoretically for one dimensional systems.

   \begin{figure}
   \includegraphics{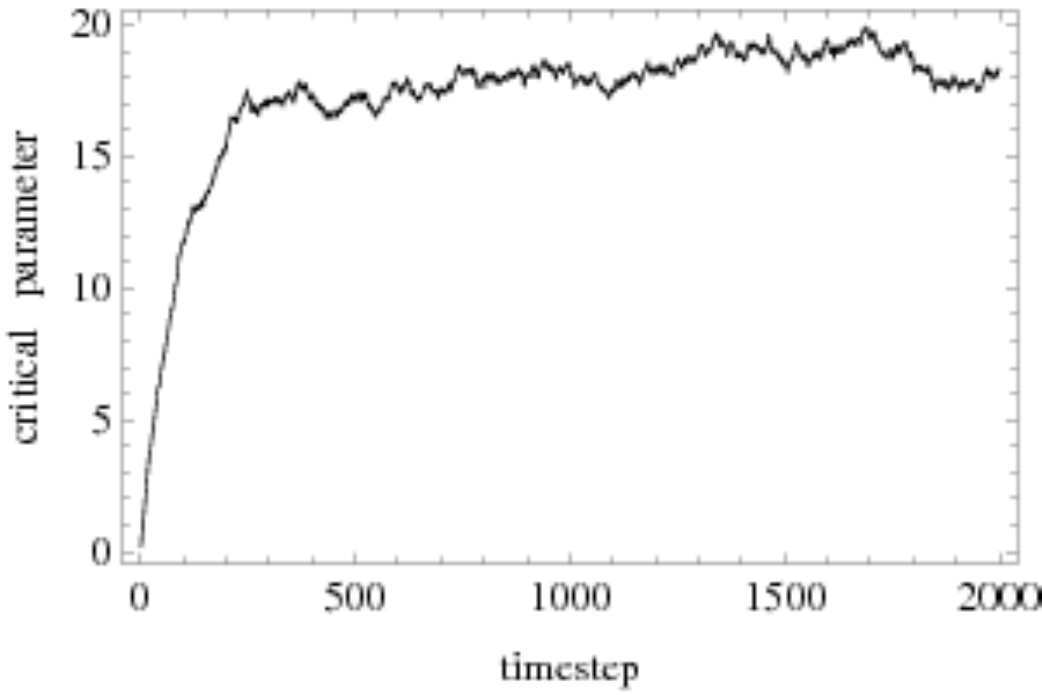}
   \caption{The evolution of the critical parameter with simulation timestep, for $\chi = 5$ and $V_{i,j} = \sqrt{j^{-1}}$.}
   \label{fig:stress}
   \end{figure}

\begin{figure}
   \includegraphics{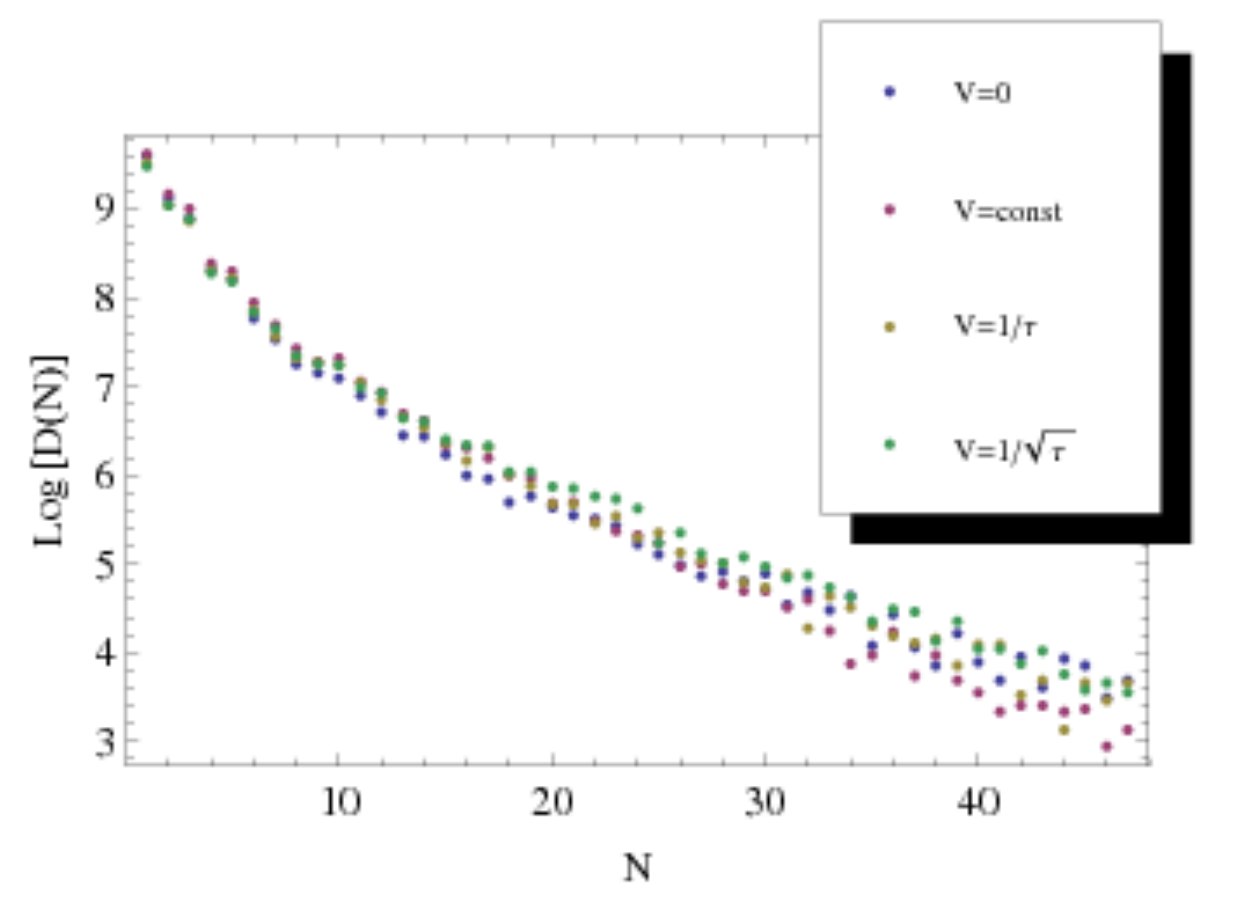}
   \includegraphics{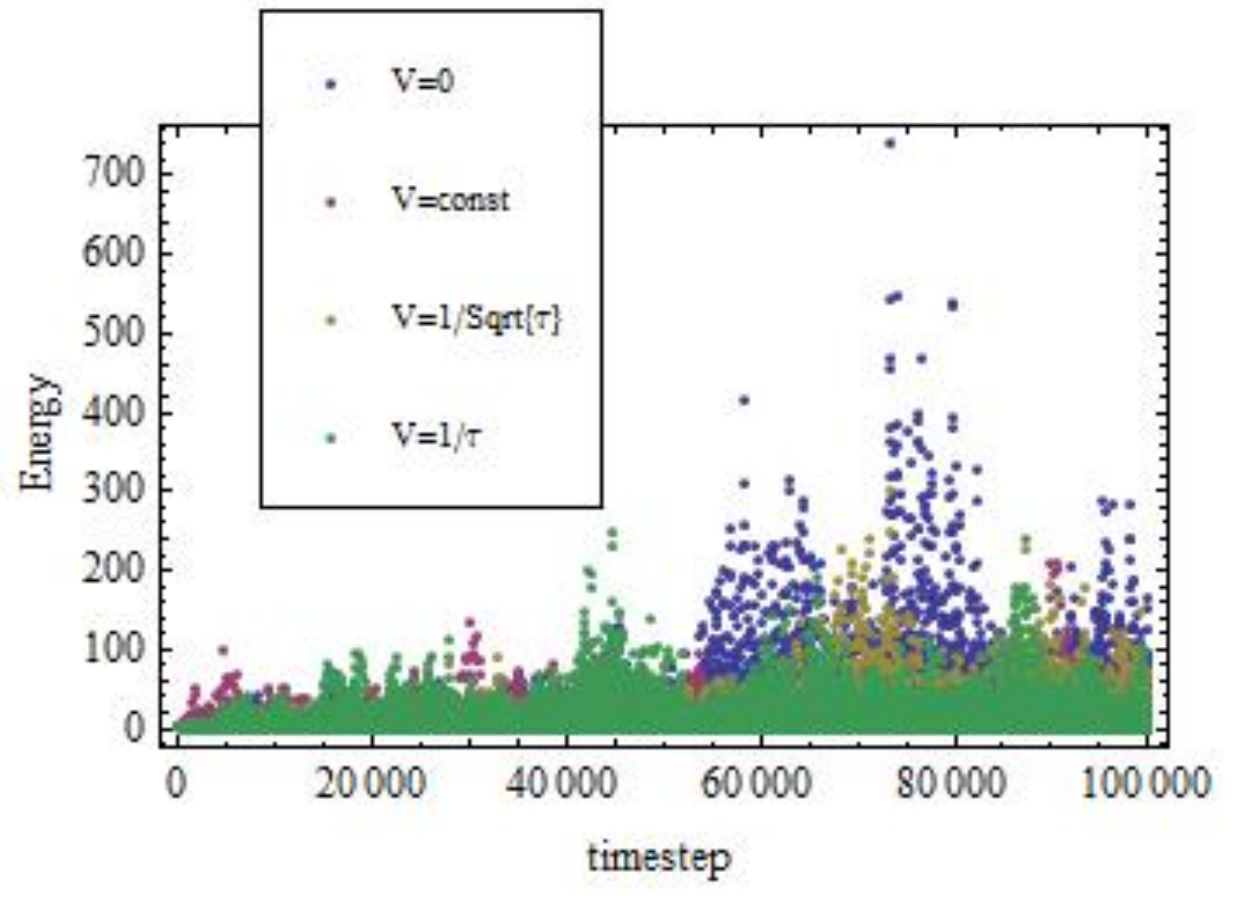}
   \caption{Left: The natural logarithm of event size distribution $D(N)$ as a function of the number of events $N$, for different velocity scaling laws, $\chi = 1$. Right: The energy release for different velocity scaling laws, for $\chi = 1$.}
   \label{fig:D(N)}
   \end{figure}

The event size distribution for different velocity profiles is shown in Fig.~\ref{fig:D(N)} for $\chi = 1$. The control parameter stabilizes over the grid (e.g. in Fig.~\ref{fig:stress}) and the magnetic field divergence is found to be under $20\%$.

%The elements which are the ingredients in "the SOC recipe" are shown in Table~\ref{tab:recipe}. In this setup SOC is not unexpected, since SOC in 3D magnetized plasma has been proven and all the correct ingredients are present in our framework.

\begin{table}\caption{Schematic display of the main characteristics a system in SOC exhibits (left column) and their illustration in our 1D CA model.}
\begin{tabular}%[!t]
{|p{0.45\linewidth}|p{0.45\linewidth}|}\hline
    SOC & 1D magnetized flow \\ \hline

    discrete space & 1D grid \\ \hline
    local interaction & one grid cell interacts with 2 other grid
cells only at criticality (diffusion) \\ \hline
    infinitely slow external drive & stochastic loading \\ \hline
    SOC occurs as a result of threshold dynamics & when the critical
parameter $g_{cr}$ is reached \\ \hline
    dissipation & boundary dissipation \\ \hline
    observables & distribution of the number of
avalanches needed to relax one perturbation \\ \hline
\end{tabular}\label{tab:recipe}
\end{table}

The choice of parameters for the simulations set $k=1/3$. This is equivalent to saying that $\eta = 1/3 \beta ^2 /\alpha$ which is the same order of magnitude as, e.g., in Eq.~(34) of~\cite{isliker1998} in terms of characteristic time-and-length scales. The value for $\chi = v_0/v_A$ sets the values of the initial velocity to that of multiples of the characteristic Alfven velocity.

Samples of energy release of the grid for different velocity scalings are shown in Fig.~\ref{fig:D(N)}, right.

\section{Application to GRB X-Ray afterglows}\label{grb}

The X-Ray afterglows occur in a strongly (a half angle of 5 degrees) collimated magnetized outflow \cite{proga2003}, with initial boundary condition given by the input flow and mass. The flow is usually modelled as a cylinder with the lower base anchored on the polar region. Magnetic
reconnection starts on the lower surface and propagates within the cylinder. We assume that this propagation occurs as a self organized critical (SOC) phenomenon (Table~\ref{tab:recipe}). According to the investigations of observational data by~\cite{StSv1996} and \cite{wang2013}, the observational signature of the data is of a system at criticality, namely of a one dimensional SOC.

We apply the simulations and results presented in the previous section to GRBs. Estimates of numerical values of the input and output parameters can now be determined by analysis of the un-scaled equations. In case of GRBs the characteristic magnetic field may be considered to be the value inferred from observations, of the order of $B_0=10^{10}T$. The dimensionless grid can be mapped to the physical grid by noticing that $\Delta z = \beta \Delta Z$ and $\Delta t = \alpha \Delta T $. The total simulation time is $t_{sim}=N_T\alpha \Delta T $. There is one more or less subtle issues here: $N_T$ can be mapped directly to a real time \textit{when no relaxation events occur in the grid}. If such relaxation events do occur, within one simulation time step, the number of real seconds increases as needed for the grid to relax. The total simulation length is $z_{sim} = N_Z \beta \Delta Z$. The correlation between the parameters and how their quantitative interpretation changes is shown in Tables~\ref{tab:params} and~\ref{tab:foralpha}.

\begin{table}\caption{Parameter correlations for current simulations. The Dimensionless column contains parameters which are set beforehand, and which generally characterise the simulation grid; the Independent parameters are those set by observations; the Dependent columns contains the parameters with an analytical dependency with respect to the dimensionless and/or independent parameters.}
\begin{tabular}%[!t]
{| p{0.27\linewidth}  | p{0.27\linewidth} | p{0.27\linewidth}|}\hline
   Dimensionless 					 	& Dependent 							& Resulting values 		\\ \hline
    $N_T=10^5$ 					     	& $\beta = \alpha v_A$					& $\beta = \alpha  \cdot 10^7\;{\rm  m}$			\\ \hline
    $N_Z=500$  						 	& $E_R \sim \frac{\beta B_0^2}{2\mu}$	& $E_R \sim \alpha \cdot 4\cdot 10^{32}\;{\rm J}$\\ \hline
    $\Delta T = \Delta Z = 1$  		    & $\Delta t = \alpha \Delta T$			& $\Delta t = \alpha \cdot s$\\ \hline
    $\chi \in \{ 1, 5, 10, 100\}$  	    & $\Delta z = \beta \Delta Z$			& $\Delta z = \alpha \cdot 10^7 m$\\ \hline
   										& $t_{sim} = N_T \alpha \Delta T$ 		& $t_{sim} =  \alpha \cdot 10^5\;{\rm s}$\\ \hline
   										& $z_{sim} = N_Z \beta \Delta Z$ 		& $z_{sim} = \alpha \cdot  5 \cdot 10^9\;{\rm m}$\\ \hline
   									
   Independent	\\ \hline					
   	$B_0=10^{10}\;{\rm T}$ (observations) 	
    &$v_A = 10^7\;{\rm m/s}$ (corresponding to $B_0$)
    &$\alpha $ \\ \hline
\end{tabular}\label{tab:params}
\end{table}

\begin{table}\caption{Illustration of how astrophysical parameters (released energy, total duration and total length) depend on the chosen characteristic timescale ($\alpha$).}
\begin{tabular}%[!t]
{| p{0.3\linewidth} | p{0.5\linewidth} |}\hline
   	$\alpha$									& Resulting values 		\\ \hline
   	$\alpha = 10^{-4}\; {\rm s}$				& $E_R \sim 4\cdot 10^{28}\; {\rm J}$ (for $e_R=1$)\\ \hline
   												& $t_{sim} =  10\; {\rm s}$\\ \hline
   												& $z_{sim} = 5\cdot 10^5\; {\rm m}$\\ \hline
   												%\\ \hline
   	$\alpha = 10^{-2}\;{\rm  s}$				& $E_R \sim 4\cdot 10^{30}\;{\rm J}$ (for $e_R=1$)\\ \hline
   												& $t_{sim} =  1000\; {\rm s}$\\ \hline
   												& $z_{sim} = 5\cdot 10^7\; {\rm m}$\\ \hline
\end{tabular}\label{tab:foralpha}
\end{table}

We propose a few parameters characterizing the GRB X-Ray afterglows to discriminate if whether or not the proposed model (and subsequent simulations) are in agreement with the observations: the time it takes the system to reach SOC, the energy released during the flaring, the slope of the event size distribution, $a$. Another commonly used parameter is the spectral slope (the slope of the power spectra) but this is shown to be in a bijective relation with $a$.

The time it takes the simulation to reach SOC depends on the initial loading of the grid. As this is generally not known, we cannot infer any quantitative result.

Since the model itself does not discuss (nor can it) the direction of emission, then isotropy and anisotropy of the radiation cannot be decided in this framework. The amount of radiation produced can be estimated based on the dimensionless energy unit (Table~\ref{tab:foralpha}).

Quantitatively, the highest dimensionless energy recorded in the simulations was $604000$ (for constant velocity with $\chi = 100$), leading to an energy estimation of at least $10^{35}$ J (for $\alpha = 10^{-2}$ s). We are not in the position to offer a strict interpretation of this result because a clear connection between one point on our plots and one count in the detector does not exist. However, we can argue that this estimate is just a minimum: assume a sampling rate of $1$ s used to obtain the observed light curve; further, assume that the detector records an integral of the emitted light curve with time as an independent variable. With $\alpha =10^{-2}$, a detector set to observe our simulated light curve would observe at least $10^{37}$ J. Also, notice that an increase of one order of magnitude in $B_0$ would lead to an increase of three orders of magnitude in the energy estimation of the simulations.

A fit of the type $D(N)\sim N^{-a}$ for the event size distribution of all the simulated lightcurves are close to the value of $a = 1$; this is not a new result but merely a consistency check, as 1D SOC is expected to produce slope unity in the event size distribution (see e.g.~\cite{aschSoc2012} for a derivation of this result). Simultaneously, this result is in agreement with results obtained from observational data by~\cite{wang2013}.

   \begin{figure}[ht]
   \includegraphics{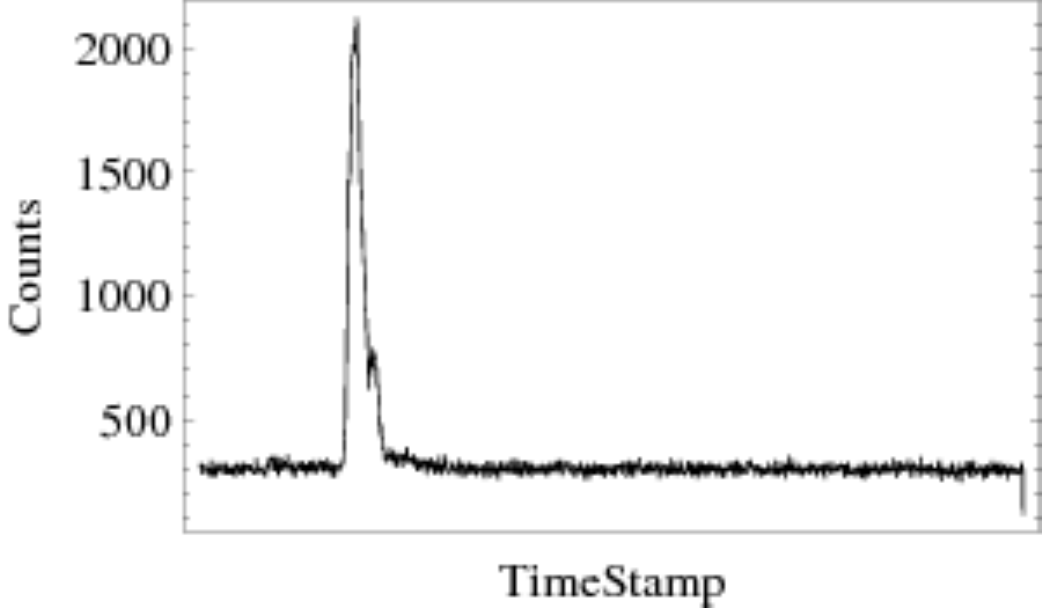}\\
   \includegraphics{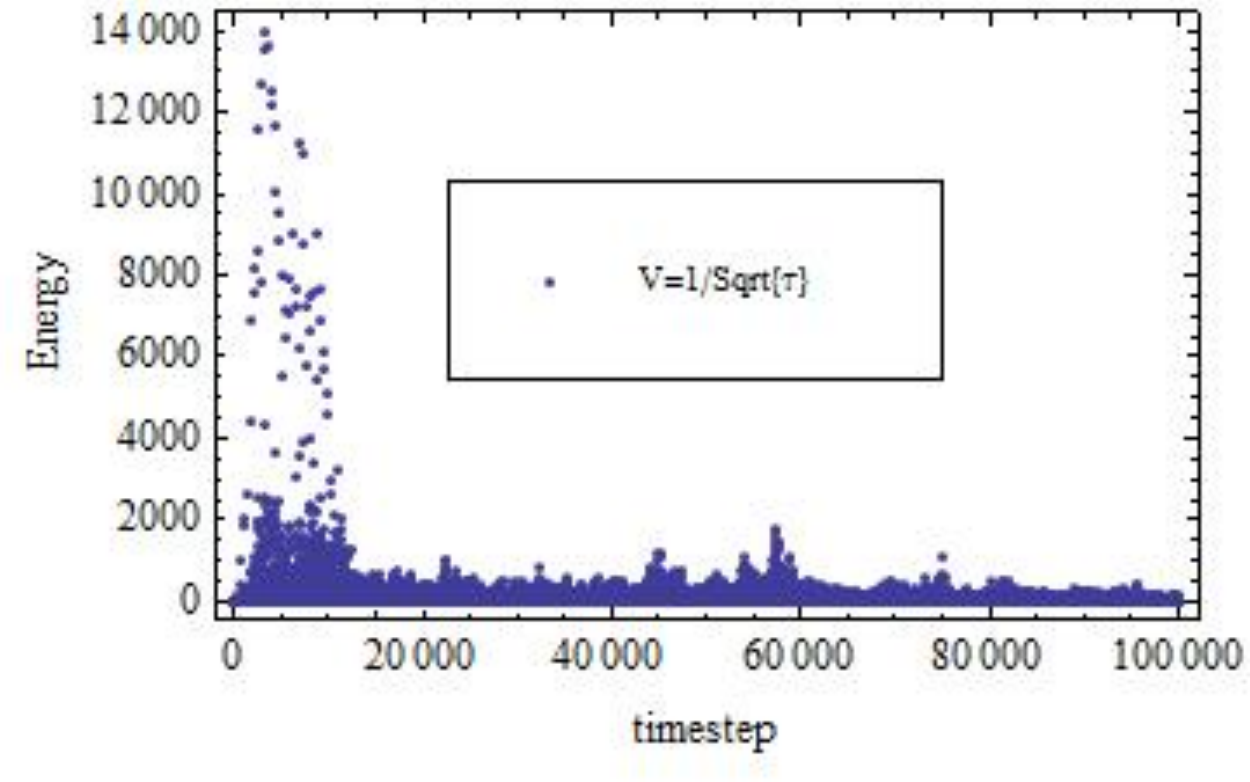}
   \caption{Upper figure: Light curve for GRB111022854 recorded with the $n_0$ Nal detector (see \cite{Collab, Fdata}). Lower figure: The energy release for a velocity scaling of $v\sim 1\sqrt{\tau}$ and $\chi = 10$.}
   \label{fig:obs}
   \end{figure}

Qualitatively, for comparison purposes, we present sample data for GRB111022854 recorded in the $n_0$ Nal detector of Fermi. We find that the bursting behaviour is well reproduced by a model with convective behaviour of the type $V\sim 1/\sqrt{\tau}$ and a large ratio of initial-to-Alfven velocity (Fig~\ref{fig:obs}).

In assessing the importance of the chosen velocity profile with respect to reproducing observational data, there are two aspects which can be discussed separately: 1) reproduction of the 1D event size distribution and 2) reproduction of observed emitted energy (light curve). As can be seen in Figure~\ref{fig:D(N)}, the model reproduces the expected event size distribution, regardless of the velocity scaling law considered. However, the bursting character observed in the light curve is reproduced only for the specific velocity scaling shown in Figure~\ref{fig:obs}.

\section{Discussions}\label{discussion}

There are several issues which must be stressed regarding this toy model. First, the model seems to show that there is a possibility that in real phenomena magnetic reconnection occurs in an one dimensional set-up. Second, the already known problem of representation in a cellular automata: have we really represented the magnetic field and its dynamics?

In real natural phenomena one dimensional magnetic reconnection is not in agreement with Maxwell's, and subsequently with the MHD, equations. One always needs at least two independent vectors in a basis to be able to describe the physics of magnetic reconnection - and in fact there exist 2D MHD setups that do this job \cite{takalo1999,vassiliadis1998,isliker1998}.

\cite{lazarian1}, \cite{lazarian2}, and \cite{lazarian3} have shown that in stochastic magnetic reconnection magnetic field line wanderings occur in a random fashion, leading to increased reconnection velocity. They calculate the ratio of the wandering length to the main direction of the magnetic field and this ratio is very small. This is how the toy model presented in this work should be regarded. Our one dimensional grid obscures the fine details of the phenomenon and follow the macroscopic evolution of the system in agreement with some microscopic laws; and this is what Cellular Automata do. We place this issue in the "{\it sweep the microscopic physics under the rug}" category.

The question of representation is what can now be called an "old" problem. In a classical three dimensional setup it has already been shown that a CA where the represented variable is the vector potential, one can claim that both qualitative and quantitative data can be extracted~\cite{isliker2000,isliker2001}. This is also valid for representing the magnetic field directly, with the only problem that $\nabla \cdot \vec{B} = 0$ cannot be controlled, but its validity can be checked a-posteriori, during the simulations.

In short, this model does not claim to do anything more than provide a CA approach to a convection plus critical diffusion equation, stemming from the set of MHD equations written for a two-dimensional configuration. This solution, at least qualitatively, fits the one dimensional Self Organized Criticality signature inferred from observations and theory.

\section{Conclusions}\label{concl}

Following the suggestion by \cite{wang2013} that X-ray flares in GRBs originate from magnetic reconnection - driven events, we have investigated the possibility of formation of Self-Organized Criticality in an one dimensional magnetized fluid flow. As mentioned in \cite{wang2013}, such a work "... could not only help to understand the central engines of GRBs, but also help to study applications of solar magnetic-reconnection theories." In  the present paper we have adopted a simplified theoretical approach for the study of magnetic reconnection, which is based on the one-dimensional form of the magnetic induction equation, with the background flow explicitly included. In the induction equation we have considered the term $-v\partial B/\partial z$ as a stochastic source term, and we have constructed an equivalent cellular automaton model to describe the evolution of the magnetized plasma flow, as well as its energy emission.

Our analysis, based on computing the event size distribution and the fact that a grid averaged value of the critical parameter stabilises to a constant value during the simulations, concludes that SOC appears in a one-dimensional magnetized setup with background flow. This result was not unexpected, as the onset of magnetic reconnection is threshold dependent and many observational and theoretical efforts, together with simulations have established that observational data can be explained by a self-organizing (dynamical) spread of reconnection events.

Even more, SOC seems to be an underlying principle of Nature. One novel pillar in this line of thought has been set by the observational analysis performed by \cite{wang2013}, and sustained by previous efforts to explain the diversity of GRBs through one single principle. This is why the possibility of the application of the present model to the GRBs afterglows was also briefly considered. Both qualitative and quantitative results show that the same grid, equipped with the same evolution law is able to produce a large diversity of light curves (e.g., Fig.~\ref{fig:D(N)}), while conserving a power-law shape for the event size distribution.

\subsection*{Acknowledgements}

We would like to thank to the anonymous referee for comments and suggestions that helped us to significantly improve our manuscript. GM is partially supported by a grant of the Romanian National Authority of Scientific Research, Program for research - Space Technology and Advanced Research - STAR, project number 72/29.11.2013.


\begin{thebibliography}{100}

\bibitem{aschSoc2012} Aschwanden, M. J., 2012, Astron. Astrophys., 539, A2
\bibitem{aschSoc2013} Aschwanden, M. J. (ed.), "Self-Organized Criticality Systems", 2013, Open Academic Press Berlin, Warsaw
\bibitem{AschOwn2013} Aschwanden M. J., "SOC Systems in Astrophysics ", 2013, 439-483, in  "Self-Organized Criticality Systems", Aschwanden, M. J. (ed.), Open Academic Press, Berlin, Warsaw
\bibitem{BaPaSp87+}  Babul A.,  Paczynski B., \&  Spergel D., 1987, Astrophys. J., 316, L49
\bibitem{bak1988} Bak P., Tang C., \& Wiesenfeld K., 1988, Physical Review A, 38, 364
\bibitem{Be10} Beloborodov  A. M., 2010, Mon. Not. R. Astron. Soc., 407, 2
\bibitem{Charbonneau} Charbonneau P., "SOC and Solar Flares" 2013, 403-437, in "Self-Organized Criticality Systems", Aschwanden, M. J. (ed.), Open Academic Press, Berlin, Warsaw
\bibitem{Chin} Chincarini G., Mao J., Margutti R., Bernardini M. G., Guidorzi C., Pasotti F., Giannios D., Della Valle M., Moretti A., Romano P., D'Avanzo P., Cusumano G., \& Giommi P., 2010, Mon. Not. R. Astron. Soc., 406, 2113
\bibitem{dimitropoulou2011}  Dimitropoulou M.,  Isliker H.,  Vlahos L., \& Georgoulis  M. K., 2011,  Astron. Astrophys., 529, A101
\bibitem{Dai} Dai Z. G., Wang X. Y., Wu X. F., \& Zhang B., 2006, Science, 311, 1127
\bibitem{gilSor1996} Gil L. \& Sornette D., 1996, Physical Review Letters, 76, 3991
\bibitem{Ha03} Harko T., 2003, Phys. Rev. D, 64, 064005
\bibitem{Ha14} Harko, T. \& Lake M. J., 2014,  Phys. Rev. D, 89, 064038 
\bibitem{isliker1998}  Isliker H.,  Anastasiadis A., Vassiliadis D.,  \& Vlahos L., 1998, Astron. Astrophys., 335, 1085
\bibitem{isliker2000} Isliker H., Anastasiadis A., \& Vlahos L., 2000, Astron. Astrophys., 363, 1134
\bibitem{isliker2001}Isliker H., Anastasiadis A., \& Vlahos L., 2001, Astron. Astrophys., 377, 1068
\bibitem{isliker2002}  Isliker H.,  Anastasiadis A., \& Vlahos L., 2002, ESASP, 506, 641I
\bibitem{lazarian1} Lazarian, A. \& Vishniac E., 2009, RevMexAA, 36, 81-88
\bibitem{lazarian2} Lazarian, A. \& Vishniac, E., 1999, Astrophys. J., 517, 700-718
\bibitem{lazarian3} Lazarian A., Vishniac E., \& Ch, J., 2004, Astrophys. J., 603, 180-197
\bibitem{lu1991} Lu E. T. \&  Hamilton R. J., 1991, Astrophys. J., 380, L89
\bibitem{lu1995} Lu E. T., 1995, Physical Review Letters, 74, 2511
\bibitem{Lyut} Lyutikov M. \& Blandford R., 2003, 	arXiv:astro-ph/0312347
\bibitem{Mar} Margutti R., Chincarini G., Granot J., Guidorzi C., Berger E., Bernardini M. G., Gehrels, N., Soderberg A. M., Stamatikos M., \& Zaninoni E., 2011, Mon. Not. R. Astron. Soc.,  417, 2144
\bibitem{MeRe93} Meszaros P. \& and Rees M. J., 1993, Astrophys. J., 405, 278
\bibitem{Me06} Meszaros P., 2006, Rep. Prog. Phys., 69, 2259
\bibitem{Metz} Metzger B. D., Giannios D., Thompson T. A., Bucciantini N., \& Quataert E., 2011, Mon. Not. R. Astron. Soc., 413, 2031
\bibitem{mineshige1994} Mineshige S., Ouchi N. B., Nishimori H., 1994, PASJ, 46, 97-105
\bibitem{mocanuAN2012} Mocanu G. \& Marcu, A., 2012, Astronomy Notes , 333, 166-173
\bibitem{moc2012} Mocanu G. \& Grumiller D., 2012, Physical Review D, 85, 105022
\bibitem{Parker} Parker, E.N., 1957, J. Geophys. Res., 62, 509-620
\bibitem{Pi92}  Piran T., 1992, Astrophys. J., 389, L45
\bibitem{Pi98} Piran T., 1998, Phys. Rept., 333, 529
\bibitem{Priest2000} Priest E. R., "Solar Magneto Hydrodynamics", 2000, D. Reidel Publishing Company, Dordrecht, Holland
\bibitem{proga2003}  Proga D.,  MacFayden A.,  Armitage P. J.,\&  Begelman M. C., 2003, ApJ, 599, L5
\bibitem{Pruessner} Pruessner, G., 2004, Studies in Self-Organized Criticality, Imperial College London
\bibitem{Re11}  Rezzolla L.,  Giacomazzo B., Baiotti L., Granot J., Kouveliotou C., \& Aloy M. A., 2011, Astrophys. J., 732, L6
\bibitem{Sornette} Sornette, D., Critical phenomena in natural sciences, Springer, 2000
\bibitem{StSv1996} Stern B. E. \& Svensson, R., 1996, Astrophys. J., 469, L109-L113
\bibitem{SuIo10} Suwa Y. \& Ioka K., 2010, Astrophys. J., 726, 107
\bibitem{Sweet} Sweet, P. A., 1958, "Magneto-Hydrostatic Equilibrium in an External Magnetic Field", in "Electromagnetic Phenomena in Cosmical Physics", Proceedings from IAU Symposium no. 6, Bo Lehnert (Ed.), International Astronomical Union, Symposium no. 6, Cambridge University Press, p. 499
\bibitem{takalo1999}  Takalo J. \& Timonen T., 1999, Geophysical Research Letters, 26, 2913
\bibitem{Collab} The Fermi-LAT Collaboration, Fermi-GBM Collaboration, Swift Collaboration, GROND Collaboration, the MOA Collaboration, 2013 ApJ 763 71
\bibitem{Fdata} The Fermi-GBM Collaboration, 2014,  http://fermi.gsfc.
    nasa.
    gov/ssc/data/analysis/scitools/gbm-grb-analysis.\\
    html
\bibitem{ToWuMe09} Toma K., Wu  X-F. \&  Meszaros P., 2009, ApJ, 707, 1404
\bibitem{vassiliadis1998} Vassiliadis D., Anastasiadis A., Georgoulis M., \& Vlahos L., 1998, Astropjysical Journal, 509, L53
\bibitem{VuBePo11} Vurm I., Beloborodov A. \& Poutanen J., 2011, ApJ, 738, 77
\bibitem{wang2013}  Wang F. Y. \& Dai Z. G., 2013, Nature Physics, 9, 465
\bibitem{Zhang08} Zhang S.N., 2008, Highlights in Astronomy, 14, 41-62
\bibitem{ZhYa10}  Zhang B. \& Yan H-R., 2011, Astrophysical Journal, 726, 90
\bibitem{Zhetal13} Zhang J., Liang E-W.,  Sun X-N., Zhang  B., Lu Y., \& Zhang S-N., 2013, ApJ, 774, L5
\end{thebibliography}
\end{document}